\documentclass[12pt,a4paper,twoside,titlepage]{article}
\usepackage{epsfig}

\newcommand{\cM}{{\cal M}}
\newcommand{\cJ}{{\cal J}}
\newcommand{\cB}{{\cal B}}

\begin{document}

\begin{titlepage}
\vspace{50mm}
\begin{center}
{\Large \bf
  A note on cactus trees:
  variational vs. recursive approach \\ 
}
\vspace{10mm}
{\large Marco Pretti}\footnote
{
  Istituto Nazionale per la Fisica della Materia (INFM)
  and Dipartimento di Fisica, Politecnico di Torino,
  Corso Duca degli Abruzzi 24, I-10129 Torino, Italy \\
  tel.: +39-011-5647373  \\
  fax: +39-011-5647399  \\
  e-mail: {\tt mpretti@athena.polito.it}
}
\end{center}
\vspace{150mm}
Submitted to the {\em Journal of Statistical Physics}
\end{titlepage}

\thispagestyle{empty}
\section*{Abstract}

In this paper we consider
the variational approach to cactus trees (Husimi trees)
and the more common recursive approach,
that are in principle equivalent for finite systems.
We discuss in detail
the conditions under which the two methods are equivalent
also in the analysis of infinite (self-similar) cactus trees,
usually investigated
to the purpose of approximating ordinary lattice systems.
Such issue is hardly ever stated in the literature.
We show (on significant test models) that
the thermodynamic quantities computed by the variational method,
when they deviates from the exact bulk properties of the cactus system,
generally provide a better approximation to the behavior
of a corresponding ordinary system.
Generalizing a property proved by Kikuchi, 
we also show that the numerical algorithm usually employed
to perform the free energy minimization in the variational approach
is always convergent. \\ \\
KEY WORDS: Cactus tree, Husimi tree, cactus approximation,
lattice model, Ising model.

\newpage
\setcounter{page}{1}
\section{Introduction}

Cactus trees are lattices with a branched topology%
~\cite{LavisBell1999,Morita1981},
and usually also a self-similar structure%
~\cite{LavisBell1999,GriffithsKaufman1982}.
Model systems on cactus trees are interesting
mainly because of two reasons.
First they often provide reliable approximations
to more realistic models on ordinary lattices%
~\cite{LavisBell1999,Gujrati1995},
and second their statistical mechanics
can be generally worked out exactly%
~\cite{LavisBell1999,Morita1981,GriffithsKaufman1982,Gujrati1995}. 
Because of these facts,
a lot of physical systems have been investigated
in the framework of tree lattices:
a variety of Ising-like models~\cite%
{Monroe1991,Monroe1992,Monroe1993,Monroe1998,%
Ananikian_et_al1998I,Ananikian_et_al1998II,%
YokoiDeOliveiraSalinas1985},
Potts models~\cite%
{Ono1986},
spin liquids~\cite%
{ChandraDoucot1994},
systems with quenched disorder~\cite%
{RiegerKirkpatrick1992}, 
polymers~\cite%
{Morita1976,StilckDeOliveira1990,StilckWheeler1992,%
BanchioSerra1995,StilckMachadoSerra1996},
abelian sandpiles~\cite%
{PapoyanShcherbakov1995/96},
electrons in binary alloys~\cite%
{MejiaLira1982}
and amorphous solids~\cite%
{Thorpe1973}.
The simplest class of lattice models, i.e. Ising models,
have been most widely investigated also on cactus trees.
In order to the approximation of systems on ordinary lattices,
it has been shown by Monroe that
the cactus approximation turns out to be particularly successful
in two relevant cases,
namely systems with multi-site interactions~\cite{Monroe1991,Monroe1992}
and frustrated systems~\cite{Monroe1998}.
In both cases the simple mean field theory
and the Bethe approximation fail
in predicting a qualitatively correct phase diagram%
~\cite{Thompson1974,Burley1972}.
In the special case of the fully frustrated
antiferromagnetic Ising model on the triangular lattice,
the same holds even for
large order Cluster Variation Method (CVM)%
~\cite{PelizzolaPretti1999},
while the cactus approximation yields
qualitatively correct results~\cite{Monroe1998}.
Recently Monroe himself has also shown that a series of
cactus approximations with larger and larger building blocks
allows not only a more precise determination of phase diagrams%
~\cite{Monroe1998}
but also quite good estimates of critical exponents%
~\cite{Monroe2001}.
As a consequence of such and other positive results, 
considerable attention has been devoted
to the properties of cactus trees%
~\cite{deMirandanetoMoraes1992/93}
and of the methods by which they are studied%
~\cite{Gujrati1995}.

In most papers dealing with cactus trees,
calculations are based on the self-similar structure of the system 
and this feature is exploited to determine its physical properties.
In the following we shall refer to such kind of treatments
as to the {\em recursive approach}.
Nevertheless it is known that cactus trees
can be studied also by means of a
{\em variational approach},
equivalent to the CVM
with a special choice of basic clusters%
~\cite{Morita1981}.
The two methods are in principle equivalent
and both give the exact solution for finite cactus trees.
Nevertheless,
to the purpose of approximating ordinary lattice systems,
one is usually interested in determining
the bulk properties of an infinite cactus tree.
This can be done exactly only by means of the recursive approach, 
by evaluating the limit of a recursion relation, 
or more precisely
by investigating the attractor of a dynamical system
defined by the recursion relation itself.
In order to employ the variational approach as well, 
one usually assumes some degree of ``translational'' invariance%
~\cite{LavisBell1999,Morita1981}.
Even when such assumption is actually verified
in the interior (bulk) of the cactus tree,
the variational free energy density evaluated in this way 
turns out to be only the bulk contribution, 
not the exact one.
The contribution of surface sites, 
whose number increases exponentially as the tree is expanded%
~\cite{LavisBell1999}, 
cannot be neglected even in the thermodynamic limit.
This has not always been stated so clearly in the literature%
~\cite{Morita1981},
and one might expect that
a minimization of the bulk free energy density
yields bulk equilibrium properties.
We point out that this is true only under certain conditions,
namely when the limit of the recursion relation exists
(that is the associated dynamical system has a fixed point,
that is the bulk invariance condition actually holds%
~\cite{LavisBell1999}), 
and as far as the equation of state is concerned.
The latter issue refers to the possibility
of multiple solutions, i.~e. coexistence phenomena, 
for which the two methods generally predict
the same solutions (if the fixed point exists) 
but different first order transitions.
In other cases (when the fixed point does not exist)
the presence of boundaries may dramatically affect
also the bulk behavior of the cactus system,
which turns out to be completely different
from that predicted by the variational approach.
By means of actual calculations
performed on significant test models
we show that,
when the variational approach deviates from the recursive approach
(that is from the exact solution),
nevertheless it generally yields a better approximation
to the behavior of a corresponding ordinary lattice system.
In the last part of the paper we discuss a particular issue
related to the variational approach.
We show an important property of the numerical algorithm 
generally used to perform the free energy minimization, 
known as Natural Iteration Method (NIM).
It turns out that
the free energy decreases at each step.
We actually generalize a proof given by Kikuchi%
~\cite{Kikuchi1974}
for the Bethe approximation,
coinciding with the variational approach to the Cayley tree%
~\cite{KurataKikuchiWatari1953}.

The paper is organized as follows.
In Sec.~II
we introduce the most important features of cactus trees.
Moreover we introduce the variational approach and the recursive approach
for the case of a finite cactus tree.
We show that the variational approach is exact
because the probability distribution
of the microscopic state of the system
has actually the factorized form
predicted by the CVM.
In Sec.~III
we consider the limit of an infinite cactus tree
with self-similar structure,
taking some restricting hypotheses,
that however include quite a large number of relevant cases.
As previously mentioned, we show that,
as far as equations of state are concerned,
the recursive approach reduces to the
variational approach,
provided the limit of the recursion relation exists.
In Sec.~IV
we work out the variational approach for three test models
already investigated by the recursive approach,
namely 
the Ising models with pure 4-spin interaction on the square cactus%
~\cite{Monroe1991},
the antiferromagnetic Ising model on the triangle cactus%
~\cite{Monroe1998},
and the Ising model with pure 3-spin interaction on the triangle cactus%
~\cite{Monroe1991}.
As previously mentioned,
we argue that the variational approach
provides a better approximation to the phase behavior
of corresponding ordinary lattice systems,
even if it generally does not coincide
with the actual bulk behavior of the cactus systems.
In Sec.~V
we prove the property of free energy decreasing
and discuss some consequences of such property
concerning the results of the previous section.
Finally in Sec.~VI
we give some concluding remarks.

\section{Fundamentals on cactus trees}

\begin{figure}
  \epsfig{file=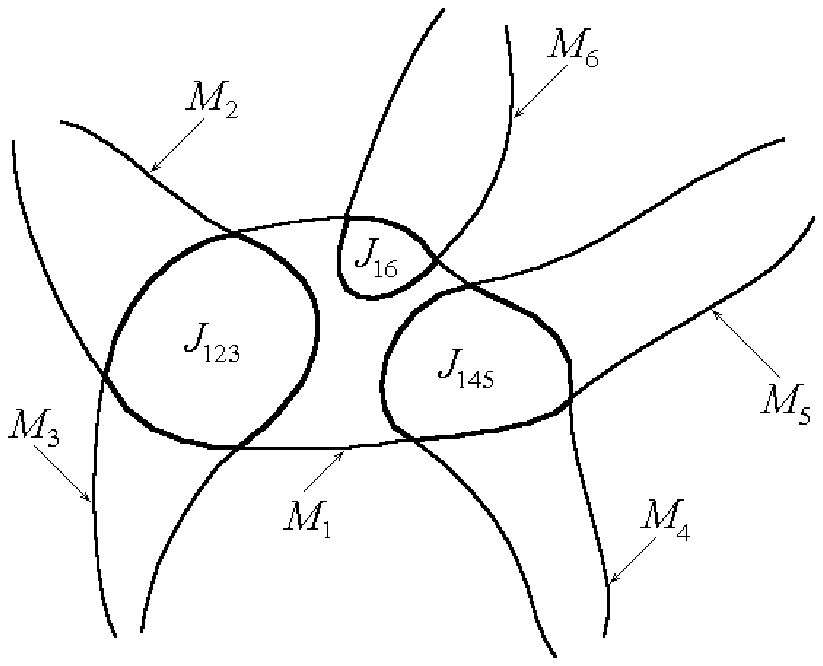,bbllx=100,bblly=550,bburx=380,bbury=760,width=10cm,clip=}
  \caption
  {
    Possible structure of a cactus tree:
    $M_i$ denote main clusters, $J_{ij\dots}$ denote joint clusters.
    The notation is such that
    $J_{ij \dots k} =
    M_i \cap M_j = \dots = M_k \cap M_i$.
    The central cluster $M_1$ contains
    $J_{123}$, $J_{145}$, and $J_{16}$. 
  }
  \label{fig:clusters}
\end{figure}

We introduce a cactus tree as a lattice with special properties.
Let us assume that
each lattice site is characterized by a state variable
and that a hamiltonian $H$ defines interactions among sites
(and of sites with external fields)
as a function of the state $X$ of the whole system.
If interactions have finite range, 
then it is possible to define a set $\cM$
of {\em main clusters} of sites,
so that the hamiltonian can be written 
as a sum over all the main clusters 
\begin{equation}
  H(X) = \sum_{M \in \cM} h_M(x_M) ,
  \label{eq:hamtot1}
\end{equation}
where $x_M$ denotes
the set of site state variables in the main cluster $M$
and $h_M(x_M)$ is a part of the hamiltonian
depending only on $x_M$.
Let us notice that the choice of $\cM$
(and also of the set of functions $\{ h_M \}_{M \in \cM}$)
may be not unique.
We require that $M \subset M'$ can never occur for any $M,M' \in \cM$;
otherwise we can give another definition of $\cM$ to avoid this.
Let us also define the set of intersections of main clusters
\begin{equation}
  \cJ \doteq
  \{ M \cap M' \left( \doteq J_{MM'} \right) \ ; \ M,M' \in \cM \} .
  \label{eq:insj}
\end{equation}
We say that a lattice is a cactus tree
if it is possible to define $\cM$ non-trivially
(i.e. not made up of a single element coinciding with the whole lattice),
so that the following property holds
for any integer $n \ge 3$ and
for any $M_1,M_2,\dots,M_n \in \cM$ 
\begin{eqnarray}
  && J_{M_1M_2}     , \dots , 
  J_{M_iM_{i+1}} , \dots , 
  J_{M_nM_1}     \ne \emptyset \ \ \Longrightarrow
\nonumber \\
  & \Longrightarrow \ \ &
  J_{M_1M_2}     = \dots =
  J_{M_iM_{i+1}} = \dots =
  J_{M_nM_1} .
  \label{eq:propj1}
\end{eqnarray}
Eq.~(\ref{eq:propj1}) states that in the lattice
there are no ``loops'' made up of main clusters,
and is the reason why the lattice can be referred to as ``tree''.
It is easy to show that such condition implies 
$J \cap J' = \emptyset$ for any $J,J' \in \cJ$.
Morita refers to clusters in the set $\cJ$ as {\em joint clusters}%
~\cite{Morita1981},
because they actually act as ``joints'' among main clusters,
and the intersection of two joint clusters is always empty.
A possible structure of a cactus tree is depicted in
Fig.~\ref{fig:clusters}.

We now introduce some more definitions and properties,
useful to determine the factorized form
of the system probability distribution (pd) $P(X)$.
Let us consider two different main clusters $M,M'$; 
we say that $M$ and $M'$ are {\em connected} if
it is possible to find $M_1,\dots,M_n \in \cM$
so that $J_{MM_1},\dots,J_{M_nM'} \ne \emptyset$.
We can assume that any $M,M' \in \cM$ are connected,
otherwise our system would be made up of two or more
non-interacting subsystems.
Moreover, given some joint cluster $J$,
we say that $M,M'$ are {\em connected excluding} $J$
if there exists a path $M_1,\dots,M_n$ connecting $M$ to $M'$
(in the sense defined previously), 
so that $J_{MM_1},\dots,J_{M_nM'} \ne J$.
We shall shortly denote this by 
$M \stackrel{J}{\sim} M'$.
It is easy to show that in this way each $J$ defines
an equivalence relation $\stackrel{J}{\sim}$ in $\cM$,
which partitions the set of main clusters into equivalence classes.
It is also possible to prove that such equivalence classes 
coincide with the {\em branches departing from} $J$,
defined as
\begin{equation}
  \cB_{JM} \doteq \{ M' \in \cM \ | \ M' \stackrel{J}{\sim} M \} .
  \label{eq:insb}
\end{equation}
for all main clusters $M \supset J$.
For example in the system of Fig.~\ref{fig:clusters}
$J_{123}$ and $J_{145}$ give rise to three branches,
while $J_{16}$ gives rise to two branches.
We now define the {\em branch hamiltonians} in the following way
\begin{equation}
  H_{JM}(x_J,X_{JM}) \doteq
  \sum_{M' \in \cB_{JM}}
  h_{M'}(x_{M'}) ,
  \label{eq:hambran1}
\end{equation}
where $X_{JM}$ denotes the state of the branch
minus the base cluster $J$.
We also define the {\em partial (branch) partition functions}
\begin{equation}
  W_{JM}(x_J) \doteq 
  \sum_{X_{JM}} e^{-\beta H_{JM}(x_J,X_{JM})} ,
  \label{eq:zbran1}
\end{equation}
where the sum runs over all possible states $X_{JM}$
and $\beta = 1/k_BT$,
being $k_B$ the Boltzmann constant
and $T$ the absolute temperature.
From Eq.~(\ref{eq:hamtot1}), Eq.~(\ref{eq:hambran1}),
and the fact that
$\{ \cB_{JM} \}_{M \supset J}$ gives a partition of $\cM$
it turns out that, for each joint cluster $J$,
the total hamiltonian can be expressed by
\begin{equation}
  H(X) =
  \sum_{M \supset J}
  H_{JM}(x_J,X_{JM}) ,
  \label{eq:hamtot2}
\end{equation}
where the sum runs over all main clusters $M$ containing $J$.
Using also Eq.~(\ref{eq:zbran1}),
this allows to write the pd $p_J(x_J)$
of the state $x_J$ of a joint cluster $J$
as follows
\begin{equation}
  p_J(x_J) = Z^{-1} \prod_{M \supset J} W_{JM}(x_J) ,
  \label{eq:pj}
\end{equation}
where $Z=\sum_Xe^{-\beta H(X)}$ is the (total) partition function.
Fig.~\ref{fig:clusters} shows that 
one more way to write the hamiltonian~(\ref{eq:hamtot1}) is,
for each main cluster $M$,
\begin{equation}
  H(X) = h_M(x_M) +
  \sum_{J \subset M}
  \sum_{\stackrel{\scriptstyle M' \supset J}{M' \ne M}}
  H_{JM'}(x_J,X_{JM'}) ,
  \label{eq:hamtot3}
\end{equation}
where the outer sum runs over all joint clusters $J$ contained in $M$,
and the inner one over all main clusters $M'$ containing $J$,
except $M$ itself.
Using again Eq.~(\ref{eq:zbran1}),
the pd $p_M(x_M)$ of the state $x_M$ of a main cluster $M$
turns out to be
\begin{equation}
  p_M(x_M) =
  Z^{-1}
  e^{-\beta h_M(x_M)}
  \prod_{J \subset M}
  \prod_{\stackrel{\scriptstyle M' \supset J}{M' \ne M}}
  W_{JM'}(x_J) .
  \label{eq:pm}
\end{equation}
We also notice that
\begin{equation}
  \sum_{M \in \cM} 1 - \sum_{J \in \cJ} (c_J-1) = 1  ,
  \label{eq:sumrule}
\end{equation}
where $c_J$ denotes the connectivity constant
of the joint cluster $J$,
i.e. the number of main clusters $M$ such that $M \supset J$.
Such ``sum rule'' can be easily proved
by considering any starting main cluster
and adding the other ones by a ``growth'' procedure.
It turns out that each joint cluster $J$
implies the addition of $c_J-1$ main cluster,
whence Eq.~(\ref{eq:sumrule}).

Taking into account
Eqs.~(\ref{eq:pj}), (\ref{eq:pm}), (\ref{eq:sumrule}),
and performing some straightforward manipulations, 
it is possible to prove that
the pd of the whole system $P(X)=Z^{-1}e^{-\beta H(X)}$ 
takes on the factorized form
\begin{equation}
  P(X) = \frac
  {\displaystyle \prod_{M \in \cM} p_M(x_M)}
  {\displaystyle \prod_{J \in \cJ} [p_J(x_J)]^{c_J-1}} .
  \label{eq:ptot}
\end{equation}
We can now write the free energy as a function of
pds of main and joint clusters only.
As far as the entropy $S$ is concerned, 
we split each contribution from a joint cluster 
among main clusters that contain it,
and write
\begin{equation}
  S/k_B = \langle - \log P(X) \rangle
  = - \sum_{M \in \cM}
  \left \langle
  \log p_M(x_M) -
  \sum_{J \subset M}
  \frac{c_J-1}{c_J} \log p_J(x_J)
  \right \rangle ,
\end{equation}
where $\langle \cdot \rangle$ denotes an ensemble average.
For the internal energy $U$,
from Eq.~(\ref{eq:hamtot1}) we simply have
\begin{equation}
  U =
  \langle H(X) \rangle = 
  \sum_{M \in \cM}
  \langle h_M(x_M) \rangle .
\end{equation}
Expanding ensemble averages, 
we can then express the free energy $F$ by
\begin{equation}
  \beta F =
  \beta U - S/k_B =
  \sum_{M \in \cM}
  \sum_{x_M} p_M(x_M)
  \varphi_M(x_M) ,
  \label{eq:bf1}
\end{equation}
where
\begin{equation}
  \varphi_M(x_M) \doteq
  \beta h_M(x_M) +
  \log p_M(x_M) -
  \sum_{J \subset M}
  b_J
  \log p_J(x_J) 
  \label{eq:fi1}
\end{equation}
and $b_J \doteq 1-1/c_J$.
It turns out that this free energy expression, which is exact,
coincides with the CVM free energy%
~\cite{An1988}
where the set of basic clusters is $\cM$.
Then the exact thermodynamic equilibrium state
can be determined by the minimization of this free energy
with respect to main cluster pds $p_M(x_M)$,
with suitable compatibility constraints for joint cluster pds $p_J(x_J)$.
The latter must be obtained as marginal distributions
for all $M \supset J$ by
\begin{equation}
  p_J(x_J) =
  \sum_{x_{M \setminus J}}
  p_M(x_M) ,
  \label{eq:margin1}
\end{equation}
where $x_{M \setminus J}$ denotes the state of
the main cluster $M$ minus the joint cluster $J$.
This concludes the general discussion
as far as the variational approach is concerned.

As far as the recursive approach is concerned,
we have to introduce a simple relation
between the partial partition function of a branch
and those of its ``sub-branches''.
Let us consider for instance a joint cluster in
Fig.~\ref{fig:clusters}
and the hamiltonian of the branch
towards the central main cluster.
It is easy to relate the latter to hamiltonians
of branches starting from the other two joint clusters displayed.
In general it is possible to write
\begin{equation}
  H_{JM}(x_J,X_{JM}) =
  h_M(x_M) +
  \sum_{\stackrel{\scriptstyle J' \subset M }{J' \ne J}}
  \sum_{\stackrel{\scriptstyle M' \supset J'}{M' \ne M}}
  H_{J'M'}(x_{J'},X_{J'M'}) ,
  \label{eq:hambran2}
\end{equation}
where the outer sum runs over
all joint clusters $J'$ contained in $M$ except $J$,
and the inner one
over all main clusters $M'$ containing $J'$,
except $M$ itself.
Hence from Eq.~(\ref{eq:zbran1})
\begin{equation}
  W_{JM}(x_J) =
  \sum_{x_{M \setminus J}}
  e^{-\beta h_M(x_M)}
  \prod_{\stackrel{\scriptstyle J' \subset M }{J' \ne J}}
  \prod_{\stackrel{\scriptstyle M' \supset J'}{M' \ne M}}
  W_{J'M'}(x_{J'}) .
  \label{eq:zbran2}
\end{equation}
By means of this equation
(in a recursive manner, starting from the boundaries)
it is possible to determine partial partition functions $W_{JM}(x_J)$
for all branches $\cB_{JM}$ departing from each joint cluster $J$
of a finite cactus tree.
Eqs.~(\ref{eq:pj}) and~(\ref{eq:pm})
then provide respectively joint and main cluster pds,
from which all equilibrium thermodynamic properties can be derived.

\section{The variational and the recursive approach to infinite cactus trees}

\begin{figure}
  \epsfig{file=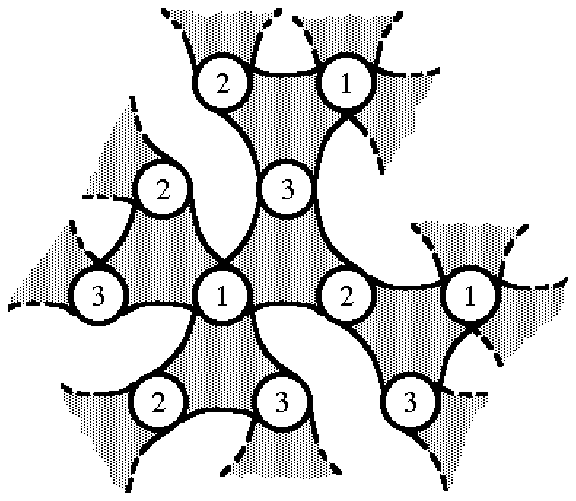,bbllx=100,bblly=590,bburx=290,bbury=750,width=10cm,clip=}
  \caption
  {
    An example of (planar) cactus tree with equivalent main clusters,
    and $n=3$ different types of joint sites.
    Connectivity constants are $c_1=3$ and $c_2=c_3=2$.
    Shaded areas denote portions of main clusters
    not covered by joint sites.
  }
  \label{fig:trcac}
\end{figure}

In this section we consider the limit of infinite cactus tree
with self-similar structure. 
This case is relevant to the approximation of
model systems on ordinary lattices,
characterized by translational invariance.
We take some restricting hypotheses,
that however include quite a large number of cases: 
we assume that
(i) site state variables are scalars, and 
(ii) joint clusters are only single sites
(we shall simply say {\em joint sites} in the following).
In order to have a self-similar structure, 
we also require that
(iii) each main cluster contains the same total number of sites
(with the same number $n$ of joint sites), and
(iv) all main cluster hamiltonians are equivalent.
The latter condition reads
\begin{equation}
  h_M(x) \equiv h(x) \ \ \ \ \ \ \forall M \in \cM  ,
  \label{eq:homh}
\end{equation}
with
\begin{equation}
  x  \doteq  \left\{ x_0,x_1,\dots,x_n \right\}.
\end{equation}
Here $x$ denotes the total state of a main cluster: 
joint site states are denoted by $x_i$ ($i=1,\dots,n$),
while $x_0$ denotes the state of
the portion of a main cluster not covered by joint sites. 
In principle we assume there are some terms in $h(x)$
that make joint sites distinguishable
(for instance interactions with $n$ different external fields).
Defining a set of connectivity constants $c_i$
($i=1,\dots,n$),
we build the cactus tree with the usual growth procedure.
We attach $c_i-1$ equivalent main clusters
to the $i$-th joint site of a starting main cluster,
and then we produce new ``generations'' iterating the procedure.
We obtain a structure of the type depicted in 
Fig.~\ref{fig:trcac},
which satisfies all previous requirements.
Let us notice that such a system turns out to be
self-similar only in the thermodynamic limit.

It is not possible to apply to this case
the variational approach as described in the previous section,
because one would have to deal with
an infinity of variational parameters.
Therefore we assume an invariance condition
\begin{equation}
  p_M(x) \equiv p(x) 
  \label{eq:homp}
\end{equation}
for main clusters $M$ in the interior (bulk) of the tree,
that is far from the surface. 
In this way the variational approach
can only determine the bulk equilibrium state,
that is $p(x)$,
which should approximate
that of a corresponding ordinary lattice model, 
and is assumed independent of boundary conditions.
In the hypothesis that Eq.~(\ref{eq:homp}) holds, 
in the bulk we shall have only a number $n$ of
(in principle) different ``types'' of joint sites,
i.e. $n$ different joint site pds $p_i(x_i)$ ($i=1,\dots,n$). 
For convenience we denote
the state of a bulk main cluster minus the $i$-th joint site
as 
\begin{equation}
  x_{\setminus i}  \doteq
  \left\{  x_0,x_1,\dots,x_{i-1},x_{i+1},\dots,x_n  \right\}  .
\end{equation}
Accordingly joint site pds can be written as
marginal distributions by 
\begin{equation}
  p_i(x_i) = \sum_{x_{\setminus i}} p(x) .
  \label{eq:margin2}
\end{equation}
As mentioned in the introduction,
we only evaluate the bulk free energy density $f$ (per main cluster)
and perform a minimization of $f$,
expecting to determine the bulk equilibrium state.
Taking into account Eqs.~(\ref{eq:bf1}),~(\ref{eq:fi1}),
and the invariance assumption Eq.~(\ref{eq:homp}),
we can write
\begin{equation}
  \beta f = \sum_{x} p(x) \varphi(x) ,
  \label{eq:bf2}
\end{equation}
where
\begin{equation}
  \varphi(x) \doteq \beta h(x) + \log p(x) - \sum_{i=1}^{n}
  b_i \log p_i(x_i) ,
  \label{eq:fi2}
\end{equation}
and $b_i \doteq 1-1/c_i$.
The free energy density $f$ is a functional in $p(x)$ only,
being $p_i(x_i)$ dependent on them via Eq.~(\ref{eq:margin2}).
We then work out the minimization with respect to $p(x)$, 
using the Lagrange multiplier method
to impose the normalization constraint
\begin{equation}
  \sum_x p(x) = 1  .
  \label{eq:normal1}
\end{equation}
Let us define the functional 
\begin{equation}
  \beta f_\lambda = \beta f - \lambda \left[ \sum_{x} p(x) - 1 \right] ,
\end{equation}
where $\lambda$ is the unknown Lagrange multiplier,
to be determined imposing the constraint.
Taking the derivatives of $f_\lambda$ with respect to $p(x)$
(making use of Eq.~(\ref{eq:margin2})), 
and setting them to zero, we obtain
\begin{equation}
  p(x) = z^{-1} e^{-\beta h(x)}
  \prod_{i=1}^{n} \left[ p_i(x_i) \right]
  ^{b_i}  ,
  \label{eq:nim}
\end{equation}
where $z$ is related to $\lambda$ in an irrelevant way.
We then simply take the sum of both sides of Eq.~(\ref{eq:nim})
over all the main cluster states $x$, 
and impose Eq.~(\ref{eq:normal1}), 
obtaining
\begin{equation}
  z = \sum_{x} e^{-\beta h(x)}
  \prod_{i=1}^{n} \left[ p_i(x_i) \right]
  ^{b_i} .
  \label{eq:z1}
\end{equation}
Eq.~(\ref{eq:nim}), with $z$ defined by Eq.~(\ref{eq:z1}),
provides a fixed point equation for $p(x)$,
which is usually solved via an iterative procedure
known as Natural Iteration Method (NIM)%
~\cite{Kikuchi1974}. 
Different solutions may be found starting the procedure
from different guesses $p(x)$,
and the stable phase is determined as the solution 
with the lowest free energy density $f$. 
The latter can be evaluated
by taking the logarithm of both sides of Eq.~(\ref{eq:nim}),
substituting into Eq.~(\ref{eq:fi2}) and then into Eq.~(\ref{eq:bf2}),
yielding
\begin{equation}
  \beta f = - \log z  ,
  \label{eq:bflogz}
\end{equation}
where $z$ has to be computed at each iteration.
In the following we shall verify that such a criterion of stability 
generally does not predict the actual first order phase transitions
for an infinite cactus system,
because it does not take into account surface contributions
to the free energy.
Nevertheless it seems to be the most reasonable way
to approximate the phase behavior of a corresponding ordinary system%
~\cite{Morita1981,Gujrati1995}.

Let us turn to the recursive method
and consider a branch of our cactus tree.
It is easy to see that 
partial partition functions $W_{JM}(x_J)$
depend only on the joint site index $i$
and the number $k$ of generations attached to it.
For a $k$-th generation branch, 
partial partition functions can then be denoted by
$W_{i,k}(x_i)$ ($i=1,\dots,n$),
and the recursion relation Eq.~(\ref{eq:zbran2}) reads
\begin{equation}
  W_{i,k}(x_i) =
  \sum_{x_{\setminus i}}
  e^{-\beta h(x)}
  \prod_{\stackrel{\scriptstyle i'=1}{i' \ne i}}^{n}
  \left[ W_{i',k-1}(x_{i'}) \right]^{c_{i'}-1}  .
  \label{eq:scm}
\end{equation}
Such equation could determine in principle
all partial partition functions
until the thermodynamic limit $\lim_{k \to \infty} W_{i,k}(x_i)$,
necessary to obtain bulk properties.
In practice this is not actually possible,
because such limit equals infinity.
However one usually performs some simple manipulations,
leading to a feasible recursion relation, 
but this will be shown in the next section for particular examples.
By now we only show explicitly that
the recursion relation Eq.~(\ref{eq:scm})
is equivalent to the NIM equations~(\ref{eq:nim}) and~(\ref{eq:z1})
in the thermodynamic limit $k \to \infty$,
in the hypothesis that such limit exists%
~\cite{LavisBell1999}.
The existence of the limit is actually equivalent
to the bulk invariance condition Eq.~(\ref{eq:homp}).
In the framework of the recursive approach
one usually computes the $i$-th (bulk) joint site pd $p_i(x_i)$
by
(i) attaching $c_i$ $k$-th generation branches of the $i$-th type,
(ii) evaluating the central site pd $p_{i,k}(x_i)$, and 
(iii) taking the limit
\begin{equation}
  p_i(x_i) = \lim_{k \to \infty} p_{i,k}(x_i)  .
  \label{eq:limp}
\end{equation}
Notice that one actually considers the central site
of $n$ (in principle) different trees.
To be rigorous,
this does not evaluate correctly
the properties of sites close to the surface of the real tree,
but becomes exact for bulk sites,
still in the hypothesis that the limit $k \to \infty$ exists.
Specializing Eq.~(\ref{eq:pj}) we can write
\begin{equation}
  p_{i,k}(x_i) = Z_{i,k}^{-1}
  \left[ W_{i,k}(x_i) \right]^{c_i}  ,
  \label{eq:pi}
\end{equation}
where
\begin{equation}
  Z_{i,k} = \sum_{x_i}
  \left[ W_{i,k}(x_i) \right]^{c_i} 
  \label{eq:zi}
\end{equation}
is the partition function of the $i$-th tree
made up of $c_i$ $k$-th generation branches.
All $Z_{i,k}$ tend to the partition function $Z$
of the infinite tree in the limit $k \to \infty$.
Substituting Eq.~(\ref{eq:pi}) into Eq.~(\ref{eq:scm}),
and multiplying both sides by $\left[ p_{ik}(x_i) \right]^{b_i}$,
we obtain
\begin{equation}
  p_{i,k}(x_i) =
  Z_{i,k}^{-1/c_i}
  \prod_{\stackrel{\scriptstyle i''=1}{i'' \ne i}}^{n}
  Z_{i'',k-1}^{b_{i''}}
  \cdot
  \sum_{x_{\setminus i}}
  e^{-\beta h(x)}
  [ p_{i,k}(x_i) ]^{b_i} 
  \prod_{\stackrel{\scriptstyle i'=1}{i' \ne i}}^{n}
  [ p_{i',k-1}(x_{i'}) ]^{b_{i'}}  .
  \label{eq:scmnim}
\end{equation}
Let us define
\begin{equation}
  z_{i,k} \doteq
  \sum_{x}
  e^{-\beta h(x)}
  [ p_{i,k}(x_i) ]^{b_i} 
  \prod_{\stackrel{\scriptstyle i'=1}{i' \ne i}}^{n}
  [ p_{i',k-1}(x_{i'}) ]^{b_{i'}}  .
  \label{eq:zik}
\end{equation}
Using Eq.~(\ref{eq:pi}),
the fact that $\sum_x \equiv \sum_{x_i} \sum_{x_{\setminus i}}$,
Eq.~(\ref{eq:scm}), and Eq.~(\ref{eq:zi}), it is possible to prove that
\begin{equation}
  z_{i,k} =
  Z_{i,k}^{1/c_i}
  \prod_{\stackrel{\scriptstyle i'=1}{i' \ne i}}^{n}
  Z_{i',k-1}^{-b_{i'}}  ,
  \label{eq:z2}
\end{equation}
which ensures normalization of $p_{i,k}(x_i)$.
Moreover it is evident from Eq.~(\ref{eq:zik}) that,
if the limit Eq.~(\ref{eq:limp}) exists, then
\begin{equation}
  \lim_{k \to \infty}
  z_{i,k} = z 
  \label{eq:limz}
\end{equation}
(independently of $i$),
where $z$ is defined by Eq.~(\ref{eq:z1}).
As a consequence,
remembering Eq.~(\ref{eq:margin2}),
we see that Eq.~(\ref{eq:scmnim}) in the limit $k \to \infty$
is equivalent to Eq.~(\ref{eq:nim}) ``marginalized'' to joint site pds
(i.e. after a summation of both sides over $x_{\setminus i}$),
which proves the equivalence with the NIM.

By the way we also notice that,
taking into account
Eqs.~(\ref{eq:bflogz}),~(\ref{eq:limz}), and~(\ref{eq:z2}), 
we can write
\begin{equation}
  \beta f =
  \lim_{k \to \infty}
  \left[
  -\frac{1}{c_i}
  \log Z_{i,k}
  +\sum_{\stackrel{\scriptstyle i'=1}{i' \ne i}}^{n}
  \frac{c_{i'}-1}{c_{i'}}
  \log Z_{i',k-1}
  \right]  .
  \label{eq:bfgujrati}
\end{equation}
When a Cayley tree is considered, 
Eq.~(\ref{eq:bfgujrati}) reduces to the formula proposed by Gujrati
(Eq.~(3) in Ref.~\cite{Gujrati1995})
to evaluate the bulk free energy density 
in the framework of the recursive approach.

\section{Test models}

In this section we consider three test models,
which we find very significant 
and have been previously investigated by the recursive approach%
~\cite{Monroe1991,Monroe1998}.
We perform variational calculations 
and compare the results with those of the recursive approach.
This should clarify the discussion of the previous section
and point out analogies and differences
between the two methods,
relating them to both
the bulk behavior of infinite cactus trees
and that of corresponding ordinary lattice systems.

The first model is an Ising model
with pure 4-spin interaction
and uniform magnetic field%
~\cite{Monroe1991}
on the square cactus
(main clusters are square ``plaquettes'' of four sites).
Each site is a joint site,
characterized by a spin state variable ($s_1,s_2,s_3,s_4 = \pm 1$),
and connectivity constants are $c_1=c_2=c_3=c_4=c$.
The main cluster hamiltonian is
\begin{equation}
  h(s_1,s_2,s_3,s_4) = -J s_1s_2s_3s_4 - H \frac{s_1+s_2+s_3+s_4}{c}  ,
  \label{eq:hsqcac}
\end{equation}
where $J>0$ is the 4-spin coupling constant
and $H$ is the magnetic field.
The NIM equations~(\ref{eq:nim}) take the form
\begin{equation}
  p(s_1,s_2,s_3,s_4) =
  z^{-1} e^{-\beta h(s_1,s_2,s_3,s_4)}
  \left[ p_1(s_1) p_2(s_2) p_3(s_3) p_4(s_4) \right]^b  ,
  \label{eq:nimsqcac}
\end{equation}
where $p(s_1,s_2,s_3,s_4)$ denotes the main cluster pd,
$b \doteq 1-1/c$,
and $z$ is determined as usual by normalization.
In principle we can distinguish four different site pds
$p_i(s_i)$ and four different magnetizations
\begin{equation}
  m_i = \langle s_i \rangle = p_i(+) - p_i(-)
  \label{eq:m}
\end{equation}
for $i=1,2,3,4$,
but from the calculation we obtain only homogeneous phases
with magnetization $m_i \equiv m$ independent of $i$.

As far as the recursive method is concerned,
on the Ising square cactus Eq.~(\ref{eq:scm}) reads
\begin{equation}
  W_{1,k}(s_1) =
  \sum_{s_2=\pm1} \sum_{s_3=\pm1} \sum_{s_4=\pm1}
  e^{-\beta h(s_1,s_2,s_3,s_4)}
  \left[ W_{2,k-1}(s_2) W_{3,k-1}(s_3) W_{4,k-1}(s_4) \right]^{c-1}
  \label{eq:scmsqcac}
\end{equation}
for site $1$, and similarly
(by a circular permutation of subscripts)
for sites $2,3,4$.
Eqs.~(\ref{eq:pi}) and~(\ref{eq:zi}) as a whole read,
for $s=\pm1$
\begin{equation}
  p_{i,k}(s) = \frac
  {\left[ W_{i,k}(s) \right]^c}
  {\left[ W_{i,k}(+) \right]^c + \left[ W_{i,k}(-) \right]^c}  .
  \label{eq:pi2}
\end{equation}
We can define the ratio
\begin{equation}
  r_k \doteq \frac{W_{i,k}(+)}{W_{i,k}(-)}
  \label{eq:r}
\end{equation}
independently of $i$,
due to the fact that one assumes homogeneous boundary conditions 
and the main cluster hamiltonian possesses a dihedral symmetry.
Finally, using Eqs.~(\ref{eq:m}),~(\ref{eq:pi2}) and~(\ref{eq:r}),
we can compute the magnetization as
\begin{equation}
  m = \lim_{k \to \infty} \frac{r_k^c-1}{r_k^c+1}  ,
\end{equation}
where, due to Eq.~(\ref{eq:scmsqcac}), $r_k$ obey the equation
\begin{equation}
  r_k = a \frac
  {a^3dr_{k-1}^{3(c-1)} + 3a^2r_{k-1}^{2(c-1)} + 3adr_{k-1}^{(c-1)} + 1}
  {a^3r_{k-1}^{3(c-1)} + 3a^2dr_{k-1}^{2(c-1)} + 3ar_{k-1}^{(c-1)} + d}  ,
  \label{eq:recsqcac}
\end{equation}
where $a \doteq e^{2\beta H/c}$ and $d \doteq e^{2\beta J}$.
Eq.~(\ref{eq:recsqcac}) is solved recursively
with the free boundary condition $r_0 = a$,
corresponding to a magnetic field equal to $H$
acting on all boundary sites.

\begin{figure}
  \epsfig{file=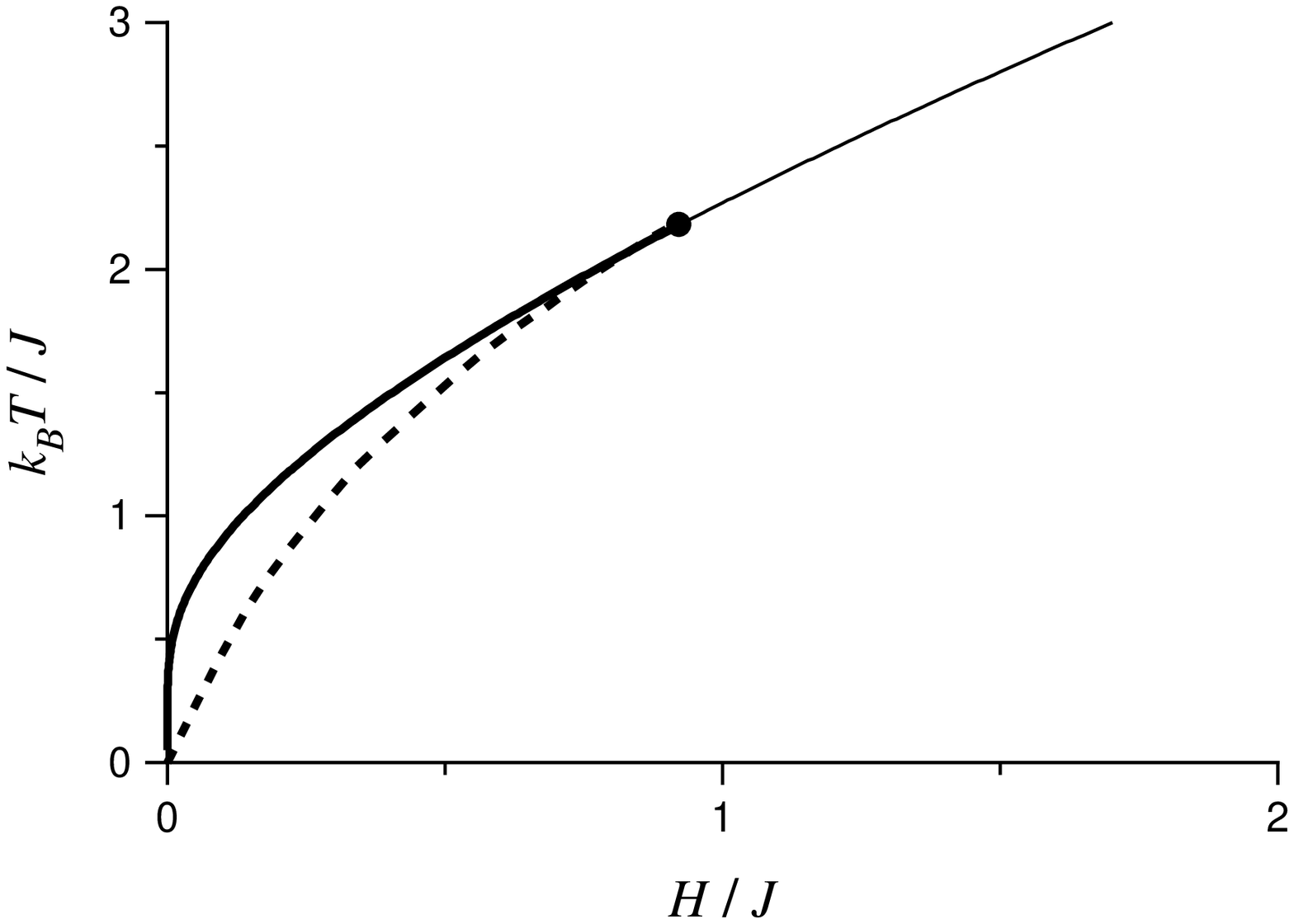,bbllx=0,bblly=340,bburx=595,bbury=740,width=10cm,clip=}
  \caption
  {
    Phase diagram of the 4-spin Ising model on the square cactus
    (temperature vs. magnetic field).
    A dashed line denotes the first order transition,
    computed by the recursive method with free boundary conditions.
    A circle denotes the critical point.
    A thick solid line denotes the same transition
    evaluated by the variational method.
    A thin solid line represents the self-dual line.
  }
  \label{fig:Monroe1991}
\end{figure}

In Fig.~\ref{fig:Monroe1991}
we report the phase diagrams obtained by both methods for $H>0$
(field inversion $H \to -H$ simply implies $m \to -m$).
We have set $c=4$,
in order to approximate the model on the ordinary square lattice.
The phase diagrams turn out to be qualitatively correct,
unlike that obtained by the mean field theory 
(see Ref.~\cite{Monroe1991} for a discussion).
We obtain a first order phase transition line at $H \ne 0$,
which separates a phase with lower magnetization
from a phase with higher magnetization. 
The line terminates at a critical point.
According to the previous section,
the equation of state is the same for both methods,
due to the fact that the recursion relation Eq.~(\ref{eq:recsqcac})
has always a fixed point.
On the contrary
the phase diagrams are only qualitatively equivalent 
but quantitatively different.
This is due to the fact that in the variational approach 
a first order transition is determined by a crossover
of the bulk free energy densities of two different phases, 
i.e. two solutions of the NIM equations 
obtained by different guess values.
On the contrary,
the recursive method has a fixed starting point,
corresponding to the boundary conditions,
and detects a first order transition as an abrupt change
(driven by model parameters)
in the attractor of the dynamical system defined by the recursion relation.
The transition observed in this way
is the actual transition for the system on the cactus tree.
Such difference is hardly ever pointed out in the literature%
~\cite{LavisBell1999,Morita1981}.
As it could be expected,
when the two competing phases degenerate into one
(i.e. at the critical point)
the two methods give the same result,
which confirms the fact that they are equivalent
as far as the equation of state is concerned.
Let us finally notice that the phase transition of the cactus system
depends on boundary conditions, 
while that predicted by the variational approach
is completely insensible to them.
This makes the latter method
not exact in predicting the phase behavior of cactus trees,
but more suitable to approximate
(translationally invariant) systems on ordinary lattices,
which is very well verified in the present case.
Let us notice that
the Ising model with pure 4-spin interaction
on the ordinary square lattice is self-dual%
~\cite{GruberHintermannMerlini1977},
and phase transitions should occur (if they do)
along the line given by
\begin{equation}
  \sinh(2 \beta J) \sinh(2 \beta H) = 1  .
\end{equation}
From Fig.~\ref{fig:Monroe1991} we observe that
the transition line obtained by the variational approach
nearly coincides (within a tolerance $\sim 10^{-3}$)
with the self-dual line.
In order to obtain better approximations
to ordinary lattice systems
also in the framework of the recursive approach,
an alternative criterion has been proposed by Monroe%
~\cite{Monroe1994}
for the location of first order transitions.
Such criterion is based on the evaluation
of the derivative of the recursion relation 
at its fixed points,
and has been verified to give the same numerical result
as the variational approach,
in the case of the Potts model.
Based on the results of Ref.~\cite{Monroe1994}
we can state that the equivalence is numerically verified
also for the present model,
because the transition line obtained by Monroe's criterion
nearly coincides with the self-dual line as well.
Nevertheless an analytic proof
of the equivalence of such criterion
with the minimization of the bulk free energy density
has not been given yet.
As previously mentioned, another criterion,
based on the evaluation of the bulk free energy density
has been proposed by Gujrati%
~\cite{Gujrati1995}
in the form of Eq.~(\ref{eq:bfgujrati}).
Unfortunately such an expression
is quite difficult to evaluate numerically
in the framework of the recursive approach,
because it involves a difference between quantities
tending to infinity in the thermodynamic limit.
On the contrary Eq.~(\ref{eq:bflogz}) shows that
the bulk free energy density comes out in a natural way
from the variational approach.

\begin{figure}
  \epsfig{file=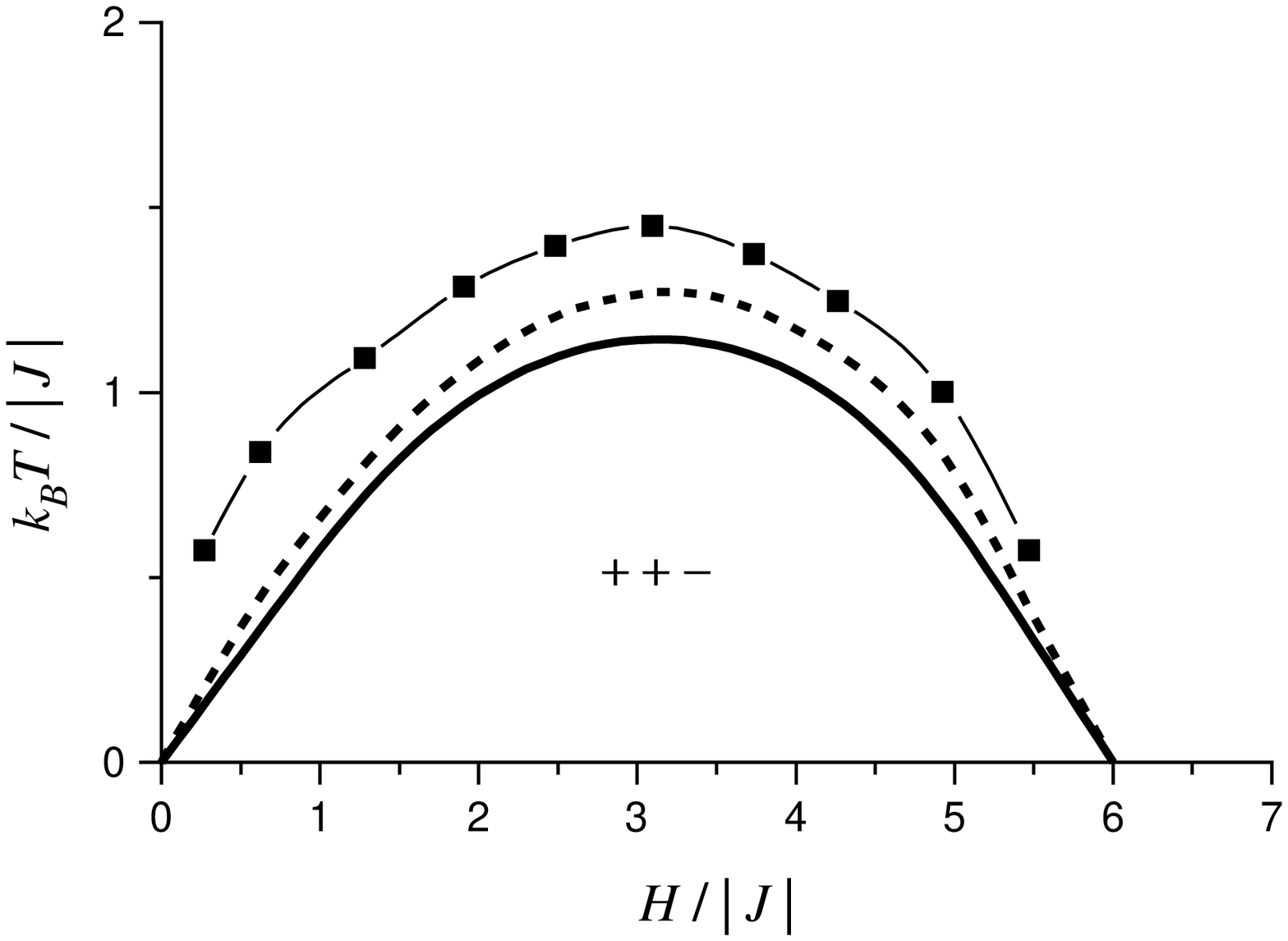,bbllx=0,bblly=340,bburx=595,bbury=740,width=10cm,clip=}
  \caption
  {
    Phase transitions of the antiferromagnetic Ising model
    on triangle cacti 
    (temperature vs. magnetic field).
    The symmetry-broken phase is denoted by $++-$.
    A thick solid line denotes the first order transition to the uniform phase,
    as predicted by the variational method.
    A dashed line denotes the same transition 
    obtained by the ``sequential'' recursive method (see the text).
    Squares denote results from Monte Carlo simulations
    of the model on the ordinary triangular lattice
    (the thin solid line is an eyeguide).
  }
  \label{fig:Monroe1998}
\end{figure}

The second test model we consider is the antiferromagnetic Ising model
with uniform magnetic field%
~\cite{Monroe1998}
on the triangle cactus 
(main clusters given by three site plaquettes).
Each site is a joint site,
characterized by a spin state variable ($s_1,s_2,s_3 = \pm 1$),
and connectivity constants are $c_1=c_2=c_3=c=3$.
The main cluster hamiltonian reads
\begin{equation}
  h(s_1,s_2,s_3) = -J (s_1s_2+s_2s_3+s_3s_1) - H \frac{s_1+s_2+s_3}{c}  ,
  \label{eq:htrcac}
\end{equation}
where $J<0$ is the antiferromagnetic coupling constant
and $H$ is the magnetic field.
The NIM equations are similar to Eq.~(\ref{eq:nimsqcac})
\begin{equation}
  p(s_1,s_2,s_3) =
  z^{-1} e^{-\beta h(s_1,s_2,s_3)}
  \left[ p_1(s_1) p_2(s_2) p_3(s_3) \right]^b  ,
  \label{eq:nimtrcac}
\end{equation}
(with obvious meaning of symbols),
while magnetizations can be obtained by Eq.~(\ref{eq:m}).
From the calculation we obtain a homogeneous phase
and a symmetry-broken phase,
where on every triangular plaquette 
we have (for $H>0$) two sites with equal positive magnetization 
and one site with negative magnetization.
The situation is inverted for $H<0$.
The phase diagram is displayed in Fig.~\ref{fig:Monroe1998}
and is symmetric with respect to $H=0$.
The transition line is always first order.
This model turns out to be interesting
as an approximation of the 
antiferromagnetic Ising model on the ordinary triangular lattice,
for which, due to frustration,
ordinary mean field like approximations%
~\cite{Burley1972},
included the CVM%
~\cite{PelizzolaPretti1999}, 
fail in predicting the (qualitatively) correct phase diagram,
and show a phase transition at zero field.

As far as the recursive method is concerned,
Eq.~(\ref{eq:scm}) reads
\begin{equation}
  W_{1,k}(s_1) =
  \sum_{s_2=\pm1} \sum_{s_3=\pm1} 
  e^{-\beta h(s_1,s_2,s_3)}
  \left[ W_{2,k-1}(s_2) W_{3,k-1}(s_3) \right]^{c-1}
  \label{eq:scmtrcac}
\end{equation}
for site $1$, and similarly
(by a circular permutation of subscripts)
for sites $2,3$.
The procedure is analogous to the previous case,
except the fact that we preserve the dependence on $i$,
in order to be able to consider inhomogeneous boundary conditions
(if homogeneous boundary conditions are imposed
the dependence on $i$ disappears
because of the dihedral symmetry of the main cluster hamiltonian).
Being
\begin{equation}
  r_{i,k} \doteq \frac{W_{i,k}(+)}{W_{i,k}(-)}  ,
  \label{eq:ri}
\end{equation}
we obtain the following recursion relation
\begin{equation}
  r_{1,k} = a \frac
  {a^2d^2r_{2,k-1}^{c-1}r_{3,k-1}^{c-1}
  + a\left(r_{2,k-1}^{c-1}+r_{3,k-1}^{c-1}\right) + 1}
  {a^2   r_{2,k-1}^{c-1}r_{3,k-1}^{c-1}
  + a\left(r_{2,k-1}^{c-1}+r_{3,k-1}^{c-1}\right) + d^2}  ,
  \label{eq:rectrcac}
\end{equation}
for site $1$, and similar ones for sites $2,3$
(circular permutation).
Magnetizations are computed by
\begin{equation}
  m_i = \lim_{k \to \infty} \frac{r_{i,k}^c-1}{r_{i,k}^c+1}  .
\end{equation}

\begin{figure}
  \epsfig{file=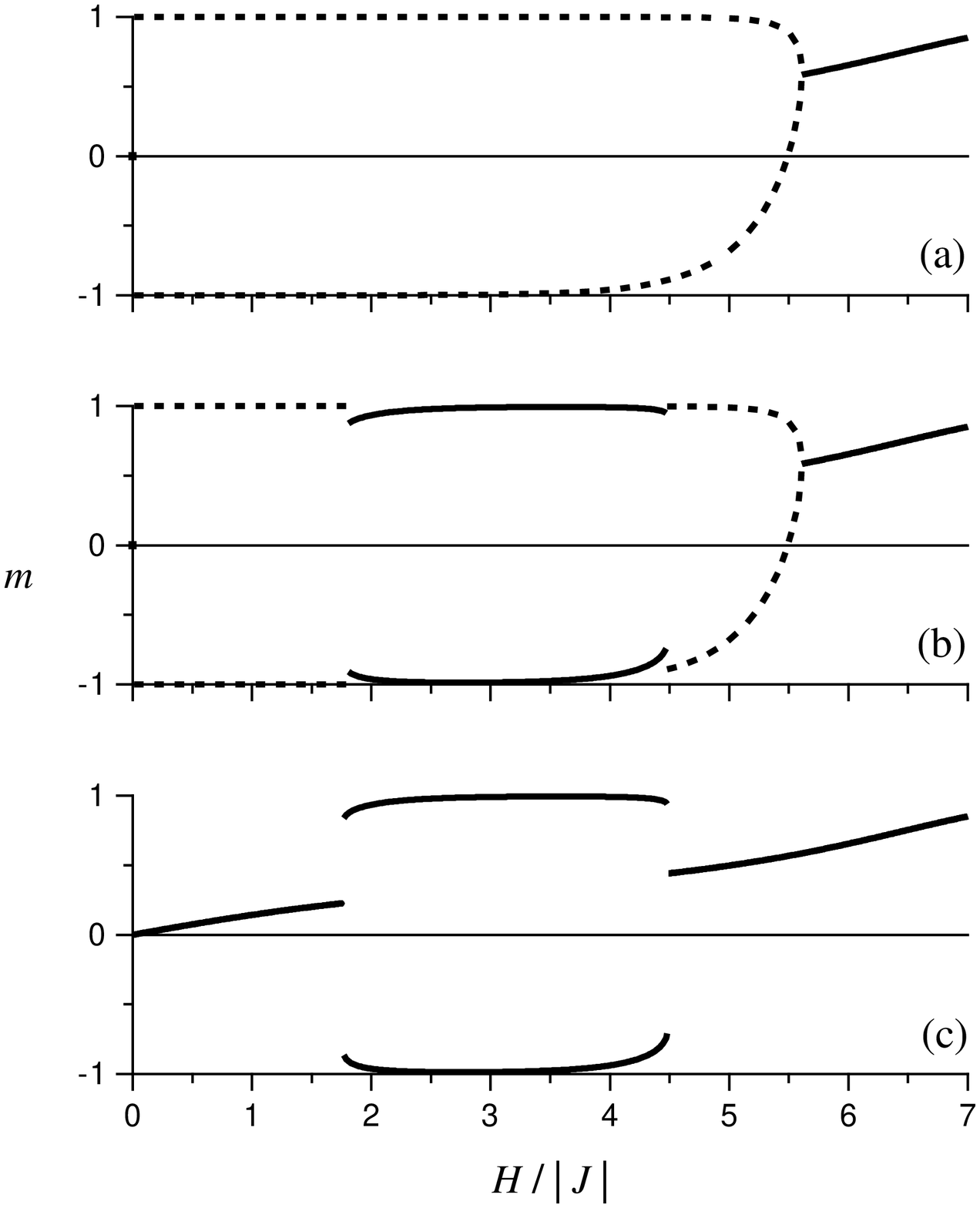,bbllx=0,bblly=70,bburx=595,bbury=740,width=10cm,clip=}
  \caption
  {
    Magnetizations of the antiferromagnetic Ising model
    on triangle cacti at fixed temperature $k_BT/|J| = 1$
    as a function of the magnetic field.
    Results obtained by the recursive method with:
    (a) free boundary conditions;
    (b) reversed field on $1/3$ of boundary sites;
    (c) free ``ragged'' boundary conditions (``sequential'' method).
    Solid lines refer to fixed point magnetizations,
    dashed lines to limit cycles (period 2).
  }
  \label{fig:mtraf}
\end{figure}

The results of the recursive approach turns out to be 
dramatically affected by boundary conditions
for the present model.
We consider a fixed temperature $k_BT/|J|=1$
and vary the field $H$,
considering the following cases.
For uniform free boundary condition
$r_{1,0} = r_{2,0} = r_{3,0} = a$ 
(magnetic field equal to $H$ on all boundary sites)
we obtain the results displayed in Fig.~\ref{fig:mtraf}(a).
The dependence on $i$ is removed but,
in a region $H \in (0,H_c)$,
the recursion relation has no fixed point 
and displays a limit cycle of period 2.
The magnetization of the central site oscillates
between the two values shown in the figure,
the positive value for even $k$ and the negative one for odd $k$.
In both cases
triangular plaquettes of consecutive generations
alternatively display two sites with positive magnetization
and one site with negative magnetization, 
or vice-versa.
On the contrary for $H=0$ and $H \ge H_c$
a fixed point exists and a paramagnetic phase
with uniform magnetization is obtained.
The latter is equivalent to that predicted by the variational approach.
We also consider the case of inhomogeneous boundary conditions
$r_{1,0} = r_{2,0} = a$ and $r_{3,0} = a^{-2}$ 
(magnetic field equal to $H$ on $2/3$ of boundary sites
and $-H$ on the remaining ones).
We obtain the results displayed in Fig.~\ref{fig:mtraf}(b).
The behavior is equivalent to the previous one,
except in a subinterval of $(0,H_c)$,
where a fixed point exists,
and the dependence on $i$ is preserved.
More precisely we obtain the same symmetry-broken phase
predicted by the variational method,
with the same numerical values of magnetizations.
This is in agreement
with the discussion performed in the previous section.
We finally compare the above results with those obtained by Monroe%
~\cite{Monroe1998}
by solving the recursion relation Eq.~(\ref{eq:rectrcac})
in a ``sequential'' way. 
Even if this actually correspond to a sligthly different system
(a tree with a ``ragged'' surface),
it turns out that a fixed point always exists,
and the behavior of magnetizations,
displayed in Fig.~\ref{fig:mtraf}(c),
is quantitatively equivalent
to that predicted by the variational approach 
(except for the positions of transitions between
the uniform phase and the symmetry-broken phase).
After the discussion of these results,
we conclude that there are some cases in which
the bulk behavior of cactus systems
does not provide reliable information
about the behavior of the corresponding
ordinary lattice system,
hence it must be employed with some caution.
Apparently contradicting ourselves, 
we have to remark that
the ``sequential'' recursive procedure worked out by Monroe gives,
with respect to the ``normal'' recursive method,
a transition line which is closer
to the Monte Carlo result~\cite{Metcalf1973} 
for the model on the ordinary triangular lattice
(see Fig.~\ref{fig:Monroe1998}).
Nevertheless this seems to be a peculiarity of the present model.

\begin{figure}
  \epsfig{file=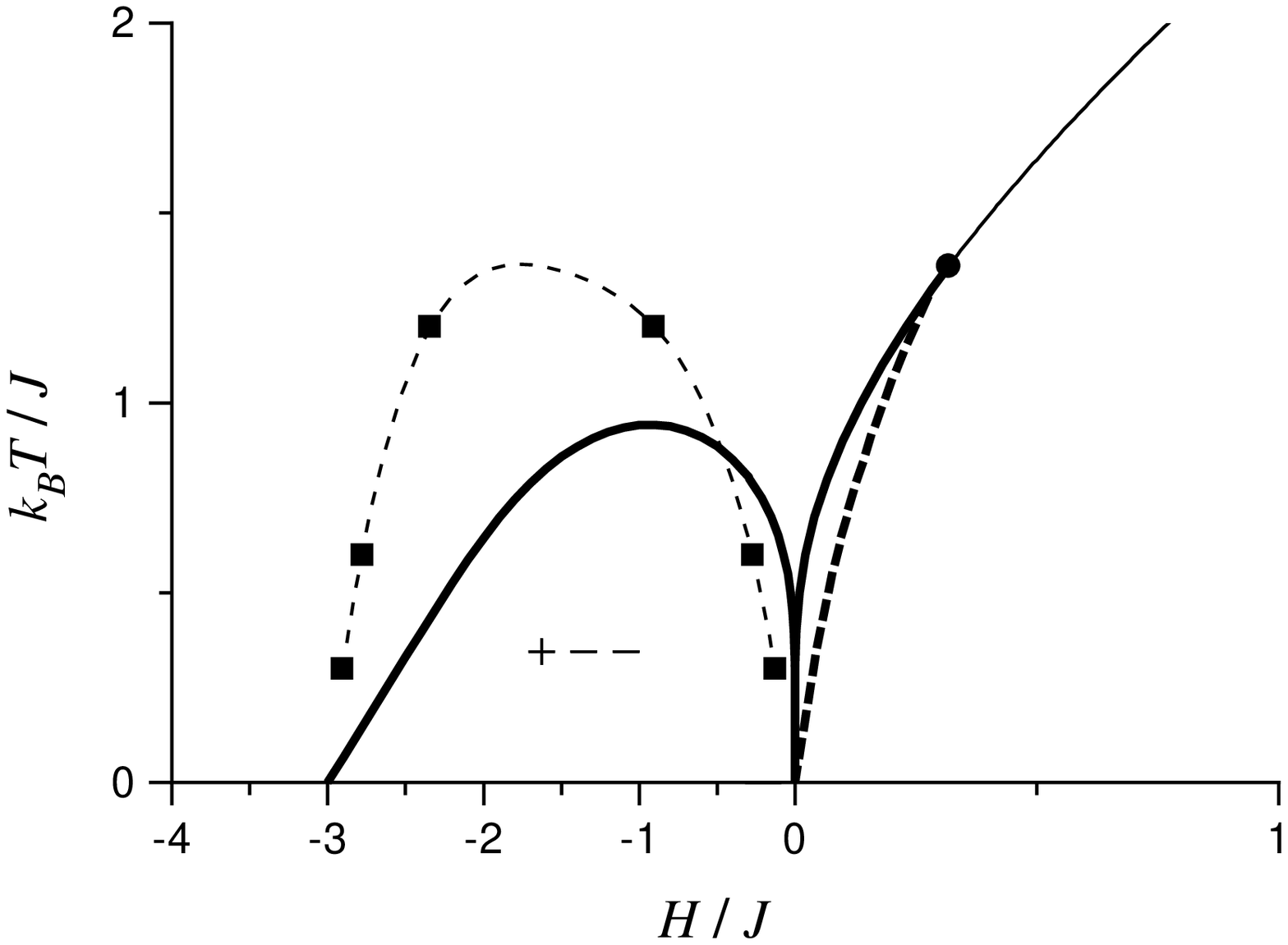,bbllx=0,bblly=340,bburx=595,bbury=740,width=10cm,clip=}
  \caption
  {
    Phase transitions of the 3-spin Ising model on the triangle cactus 
    (temperature vs. magnetic field).
    Thick solid lines denote first order transitions
    computed by the variational method.
    A circle denotes the critical point.
    A thin solid line represents the self-dual line.
    A thick dashed line denotes
    the first order transition between homogeneous phases
    obtained by the recursive method
    with free boundary conditions.
    The symmetry-broken phase region is denoted by $+--$.
    Squares mark the boundary of the region in which
    the recursive method displays limit cycles and chaos
    (the thin dashed line is an eyeguide).
  }
  \label{fig:tria}
\end{figure}

The third test model we investigate suggests that
the recursive method alone may lead to incorrect conclusions
about the physics of the model on ordinary lattice,
if no other information is available. 
We consider an Ising-like model with pure three-spin interaction
and uniform magnetic field
on the triangle cactus (introduced above).
This model has been previously investigated
by the recursive method%
~\cite{Monroe1991}, 
with the aim of approximating a model with three-spin interaction
on upward pointing (or downward pointing) triangles
of an ordinary triangular lattice.
The main cluster hamiltonian reads
\begin{equation}
  h(s_1,s_2,s_3) = -J s_1s_2s_3 - H \frac{s_1+s_2+s_3}{c}  ,
\end{equation}
where $J>0$ is the plaquette interaction, 
$H$ is the magnetic field,
and $c=3$ .
All calculations are analogous to the previous model.
The recursion relation turns out to be
\begin{equation}
  r_{1,k} = a \frac
  {a^2dr_{2,k-1}^{c-1}r_{3,k-1}^{c-1}
  + a \left(r_{2,k-1}^{c-1}+r_{3,k-1}^{c-1}\right) + d}
  {a^2 r_{2,k-1}^{c-1}r_{3,k-1}^{c-1}
  + ad\left(r_{2,k-1}^{c-1}+r_{3,k-1}^{c-1}\right) + 1} 
\end{equation}
for site $1$, and similar ones can be derived 
(by the usual circular permutation)
for sites $2,3$.
We obtain the phase diagram shown in Fig.~\ref{fig:tria}.
For $H>0$ only homogeneous phases are obtained,
with a first order transition line
which separates a phase with lower magnetization
from a phase with higher magnetization. 
The line terminates at a critical point.
The phase behavior is qualitatively similar
to the square cactus model described previously.
Similarly the two methods predict
different first order transition lines
but the same critical point.
On the contrary for $H<0$ a symmetry-broken phase appears.
According to the variational method, 
each triangular plaquette has
two sites with equal negative magnetization 
and a site with positive magnetization.
This phase is separated from the paramagnetic phase
by a first order transition line.
In almost the same region of the phase diagram
the recursive method displays a peculiar behavior,
involving limit cycles of high order
and transitions to chaos~\cite{Monroe1991}.
In Fig.~\ref{fig:tria} we report the boundaries of such region,
drawn from data published in Ref.~\cite{Monroe1991}.
We conjecture that in this case
the correct phase diagram of the ordinary lattice model
(3-spin interaction on upward/downward triangles)
is that predicted by the variational approach
(cactus approximation),
while the anomalous behavior observed
in Ref.~\cite{Monroe1991}
(and also Ref.~\cite{Ananikian_et_al1998II}
for a similar model)
is a peculiarity of the cactus tree.
Our conjecture is also supported by the fact that,
applying Monroe's sequential procedure%
~\cite{Monroe1998}
to this case
(having some analogies with the previous one),
we have obtained results
in agreement with the variational approach.

\section{Convergence of the NIM}

In this section we examine an important property
of the NIM equations~(\ref{eq:nim}).
As previously mentioned,
the NIM is a numerical iterative minimization 
of the variational free energy density.
By a generalization of the proof given by Kikuchi for the Bethe lattice%
~\cite{Kikuchi1974},
it turns out that the free energy decreases at each iteration,
and the algorithm is always convergent.
Let us give the proof
and then discuss some consequences.
Starting from Eq.~(\ref{eq:bf2}),
we write the difference between the free energies
of two consecutive steps of the iterative procedure as 
\begin{equation}
  \beta (\hat{f}-f) = \sum_{x}
  \left[ \hat{p}(x) \hat{\varphi}(x) - p(x) \varphi(x) \right]  ,
  \label{eq:deltafi1}
\end{equation}
where a hat denotes the latter step, and accordingly
\begin{equation}
  \hat{\varphi}(x) = \beta h(x) + \log \hat{p}(x) - \sum_{i=1}^{n}
  b_i \log \hat{p}_i(x_i)  ,
  \label{eq:ficap}
\end{equation}
while $\varphi(x)$ is defined by Eq.~(\ref{eq:fi2}).
Taking the the logarithm of both sides of Eq.~(\ref{eq:nim})
(where the left side is now denoted by a hat),
we can write the NIM equations in two different ways,
that are
\begin{equation}
  \log \hat{p}(x) = - \log z - \beta h(x) + \sum_{i=1}^{n}
  b_i \log p_i(x_i) 
\end{equation}
and
\begin{equation}
  \sum_{i=1}^{n} b_i \log p_i(x_i) = 
  \log z + \beta h(x) + \log \hat{p}(x)  .
\end{equation}
We substitute the former into $\hat{\varphi}(x)$, 
the latter into $\varphi(x)$,
and finally both into Eq.~(\ref{eq:deltafi1}),
yielding
\begin{equation}
  \beta (\hat{f}-f) =
  \sum_{x} p(x) \log \frac{\hat{p}(x)}{p(x)} +
  \sum_{i=1}^{n} b_i
  \sum_{x_i} \hat{p}_i(x_i) \log \frac{p_i(x_i)}{\hat{p}_i(x_i)}  .
  \label{eq:deltafi2}
\end{equation}
Let us now consider the inequality $\log \xi \le \xi-1$,
that holds for all real numbers $\xi$,
and observe that the equality holds only if $\xi=1$.
By applying this argument to both logarithms in Eq.~(\ref{eq:deltafi2}) 
and remembering that all pds are normalized at each step, 
we can finally write
\begin{eqnarray}
  \hat{f}-f & \le & 0 
  \label{eq:deltaflt0}
  \\
  \hat{f}-f & =   & 0
  \ \ \  \Longleftrightarrow \ \ \ 
  \hat{p}(x) = p(x) \ \  \forall x  .
  \label{eq:deltafeq0iff}
\end{eqnarray}
Eq.~(\ref{eq:deltaflt0}) means that
the free energy can be decreasing or constant during the procedure,
while Eq.~(\ref{eq:deltafeq0iff}) assures that
it is constant
only if the procedure has already reached convergence
(i.e. the free energy  can only decrease during the procedure).

The above property,
generally desirable for a numerical method
that aims to minimize a function, 
has some relevant consequences
and makes a significant difference
between the NIM and the recursive method.
It is evident that 
Eqs.~(\ref{eq:deltaflt0}) and~(\ref{eq:deltafeq0iff}) 
prevent the dynamical system defined by the NIM equations
from having limit cycles.
In this way the variational approach always determines
the best solution compatible with the invariance condition
Eq.~(\ref{eq:homp}), 
but cannot detect
whether such hypothesis is too restrictive or not.
The variational approach describes correctly
a symmetry breaking with a period less than or equal
to the width of a main cluster, 
but would not be able to indicate
the existence of phases with higher periodicity 
(or even incommensurate phases,
as observed for instance in the ANNNI model%
~\cite{BakVonBoehm1980}).
We then conclude that also the variational approach
must be used with some caution
in the approximation of ordinary lattice systems,
if there are reasons to suspect
that a violation of Eq.~(\ref{eq:homp}) occurs,
not only in the cactus system.
In such cases
the recursive approach is essential%
~\cite{YokoiDeOliveiraSalinas1985} 
because,
as we pointed out in the previous section, 
the dynamical system defined by the recursion relation is able,
through the features of its attractor,
to indicate the nature of the symmetry breaking.

\section{Conclusions}

In this paper we have discussed several properties
of the variational approach to cactus trees,
performing comparisons
with the more usually employed recursive approach.
First of all we have pointed out that
the variational approach is based
on an exact factorization
of the state probability distribution,
and can in principle solve exactly
finite cactus trees,
as well as the recursive method.
Moreover we have considered different issues,
concerning the bulk behavior of infinite (self-similar) cactus trees
and the approximation of ordinary lattice systems.

We have shown that the variational method
allows a simple evaluation of the bulk free energy density. 
The minimization of bulk free energy
yields the correct equation of state
for the interior of the cactus tree.
In presence of multiple solutions, i.e. coexistence phenomena,
the first order transitions determined by the variational method 
are not the exact ones for the cactus trees,
but turn out to be independent of boundary conditions,
and provide a reliable approximation (cactus approximation)
to phase transitions of corresponding ordinary systems.
On the contrary the recursive method determines
the exact bulk behavior of infinite cactus trees, 
on the basis of a change in the attractor of a dynamical system,
driven by model parameters.
Unfortunately
such behavior is strongly dependent on boundary conditions,
and usually provides poorer approximations to ordinary systems,
or even incorrect results.

We have also considered the convergence property of the algorithm
generally used to perform the free energy minimization
in the variational approach.
From the point of view of numerical analysis,
we deal with a simple iterative solution
of a set of fixed point equations,
actually the same kind of problem
which is solved in the framework of the recursive approach.
Nevertheless the peculiar form of the variational equations
allows to prove that the free energy
corresponding to the current thermodynamic state
decreases at each iteration
and remains constant only if the algorithm has already reached convergence.
We have shown on a counterexample that
this is not always true for the recursive method.
Of course the free energy decrease is a nice property in most cases,
because we are interested in free energy minima.
Nevertheless,
on the basis of results obtained on a test model,
we have suggested that it may also lead to missing some complex physics,
coming out in the form of limit cycles and/or chaotic attractors
in the recursive approach.

In the whole paper,
sometimes only by recollecting already known results, 
we have tried to give
a unified picture of the two different approaches,
pointing out analogies and differences.

\newpage

\newpage

\end{document}